\documentclass[twocolumn,preprintnumbers,amsmath,amssymb,prl]{revtex4} 
\usepackage{amsmath} 
\usepackage[utf8]{inputenc} 
\usepackage[english]{babel} 
\usepackage{graphicx} 
\usepackage{listings} 
\usepackage{soul}
\usepackage{color}

\begin{document}

\title{Crystal-to-crystal transition of ultrasoft colloids under shear}

\author{J. Ruiz-Franco$^{1}$, J. Marakis$^{2,3}$, N. Gnan$^{1,4}$, J. Kohlbrecher$^5$, M. Gauthier$^6$, M. P. Lettinga$^{7,8}$, D. Vlassopoulos$^{2,3}$, E. Zaccarelli$^{1,4}$}
\address{$^1$Dip. Di Fisica, Sapienza Universit\`a di Roma, P.le A. Moro 5, 00185 Rome, Italy}
\address{$^2$Foundation for Research and Technology-Hellas (FORTH), Institute of Electronic Structure and Laser (IESL), 100 N. Plastira Str., GR-70013 Heraklion, Greece}
\address{ $^3$University of Crete, Department of Materials Science and Technology, P.O.Box 2208, GR-71003 Heraklion, Greece}
\address{ $^4$CNR Institute of Complex Systems, Uos Sapienza, Rome, Italy}
\address{$^5$ Laboratory for Neutron Scattering, ETH Zurich \& Paul Scherrer Institut, 5232 Villigen PSI, Switzerland}
\address{$^6$ University of Waterloo, Department of Chemistry, Waterloo, ON N2L 3G1, Canada}
\address{$^7$Laboratory for Soft Matter and Biophysics, KU Leuven, Celestijnenlaan 200D, B-3001 Leuven, Belgium}
\address{$^8$Institute of Complex Systems (ICS-3), Forschungszentrum J\"ulich, 52425 J\"ulich, Germany.}
\date{\today}

\begin{abstract}

Ultrasoft colloids typically do not spontaneously crystallize, but rather vitrify, at high concentrations. Combining in-situ rheo-SANS experiments and numerical simulations we show that shear facilitates crystallization of colloidal star polymers in the vicinity
of their glass transition. With increasing shear rate well beyond rheological yielding, a transition is found from an initial bcc-dominated structure to an fcc-dominated one. This crystal-to-crystal transition is not accompanied by intermediate melting but occurs via a sudden reorganization of the crystal structure.
 Our results provide a new avenue to tailor colloidal crystallization and crystal-to-crystal transition at molecular level by coupling softness and shear.
\end{abstract}

\maketitle

Concentrated suspensions of Brownian spheres are known to undergo crystallization and/or glass transition, depending on their size  polydispersity and interaction potential\cite{pusey1991colloidal}.  With respect to hard spheres, quiescent crystallization in dense suspensions of soft colloids is in general more complicated, due to shape fluctuations and adjustment\cite{zhang2014direct,witten1986macrocrystal}. Whereas microgel-based particles crystallize at roughly the same packing fraction as hard spheres\cite{senff1999temperature,lyon2004microgel,lu2011thermosensitive}, hairy particles may do so at larger concentrations, depending of the relative core-to-grafted arm size ratio or the rate of arm exchange in the case of micelles\cite{mcconnell1993disorder, puaud2013dynamic,watzlawek1999phase, hamley2000effect}. Particles with small cores and long hairs, such as star polymers, cannot crystallize easily because arm fluctuations delay this process\cite{stiakakis2010slow,rissanou2006temperature}, despite the opposite expectations due to enhanced osmotic pressure\cite{witten1986macrocrystal}.  The slowdown of the nucleation process can also be attributed to interpenetration and clustering, which may act as an effective polydispersity suppressing crystallization\cite{gupta2015dynamic}.  It is therefore common for soft colloids to become kinetically trapped in metastable states\cite{foffi2003structural}. Colloidal glasses may crystallize eventually over time, i.e., thermodynamic equilibrium is reached, irrespectively of softness\cite{stiakakis2010slow,zaccarelli2009crystallization,rissanou2006temperature}.  
The action of an external stimulus, such as shear flow, can promote either formation or melting of ordered states, depending on its rate and strength\cite{lowen2001colloidal,vermant2005flow}.  Hence, the delicate interplay between interparticle forces and hydrodynamic interactions provides the conditions for achieving and tuning colloidal  crystallization or dynamic arrest\cite{lowen2001colloidal,vermant2005flow,holmqvist2005crystallization,imhof1994shear,besseling2012oscillatory,ackerson1988shear,ackerson1990shear,duff2007shear,koumakis2008effects,mortensen2002shear,wu2009melting}. 

The ability of shear to induce crystal formation in soft colloids is significant and well-documented \cite{jiang2007shear,slawecki1998shear,lopez2015layering,molino1998identification,mcconnell1995long,nikoubashman2012flow}. Depending on the rate of applied oscillatory or steady shear, a rich variety of crystal phases can be formed, which are often able to sustain large deformations\cite{slawecki1998shear,lopez2015layering,molino1998identification,mcconnell1995long,nikoubashman2012flow}. 
However, promoting crystallization in sheared glassy or jammed systems is challenging since their original microstructures are non-equilibrium states that may undergo phase or layering transitions while deformation of soft particles is possible\cite{stokes2008rheology,PhysRevFluids.2.093301,vlassopoulos2014tunable}.  Ultrasoft colloidal stars, for which the size and number of arms determine the 
interactions between particles\cite{watzlawek1999phase,likos1998star}, display a very rich glassy phenomenology\cite{helgeson2007viscoelasticity,vlassopoulos2014tunable}.
 At the level of particle microstructure, the interpenetration of the arms is primarily responsible for their complex rheological behavior\cite{vlassopoulos2001multiarm,kapnistos1999viscoelastic,helgeson2007viscoelasticity}, implying that shear could promote crystal formation of stars via their cooperative rearrangement which is mediated by arm disengagement. This avenue to crystallization  for hairy ultrasoft  colloids is yet to be explored.

In addition to shear-induced order, order-to-order transitions under the influence of an external stimulus are ubiquitous in colloidal systems. In particular, microgels have been found to undergo a crystal-to-crystal transition upon changing temperature in equilibrium\cite{peng2015two} and in the presence of an electric field\cite{mohanty2015multiple}. 
For block copolymer micelles\cite{mcconnell1995long,jiang2007shear,eiser2000nonhomogeneous,lopez2015layering} and microgel dumbbells\cite{chu2015colloidal} such transitions have been observed with increasing shear rate. It has been argued that the crystal-to-crystal transition occurs via two-step transformations, accompanied by the formation of an intermediate fluid phase\cite{peng2015two,lopez2015layering,chu2015colloidal}, which favors local rearrangements and subsequent recrystallization. Such intermediate melting was thus suggested to be a generic mechanism for the occurrence of crystal-to-crystal transitions\cite{sanz2015crystal}. However, whether such a transition and mechanism hold for ultrasoft colloids is an important open question.

In this Letter we investigate the consequences of an imposed shear flow on the crystallization of colloidal stars in the vicinity of their glass transition by means of in-situ rheo-SANS experiments and molecular dynamics simulations. We find that shear promotes crystallization, both under oscillatory (experiments) and steady (simulations) conditions. Moreover, we provide unambiguous evidence of a crystal-to-crystal transition under shear. 
Results from measured and calculated diffraction patterns, which are in good agreement, suggest a two-step process. At first the fluid forms a bcc-like crystal ($1^{st}$ step), which later transforms into a fcc-like one ($2^{nd}$ step) through a sudden change in the crystal structure.  Differently from previous observations\cite{peng2015two,lopez2015layering,chu2015colloidal}, we do not find evidence of an intermediate liquid phase between the two crystals.

\begin{figure}
\centering \includegraphics[width=8.0cm]{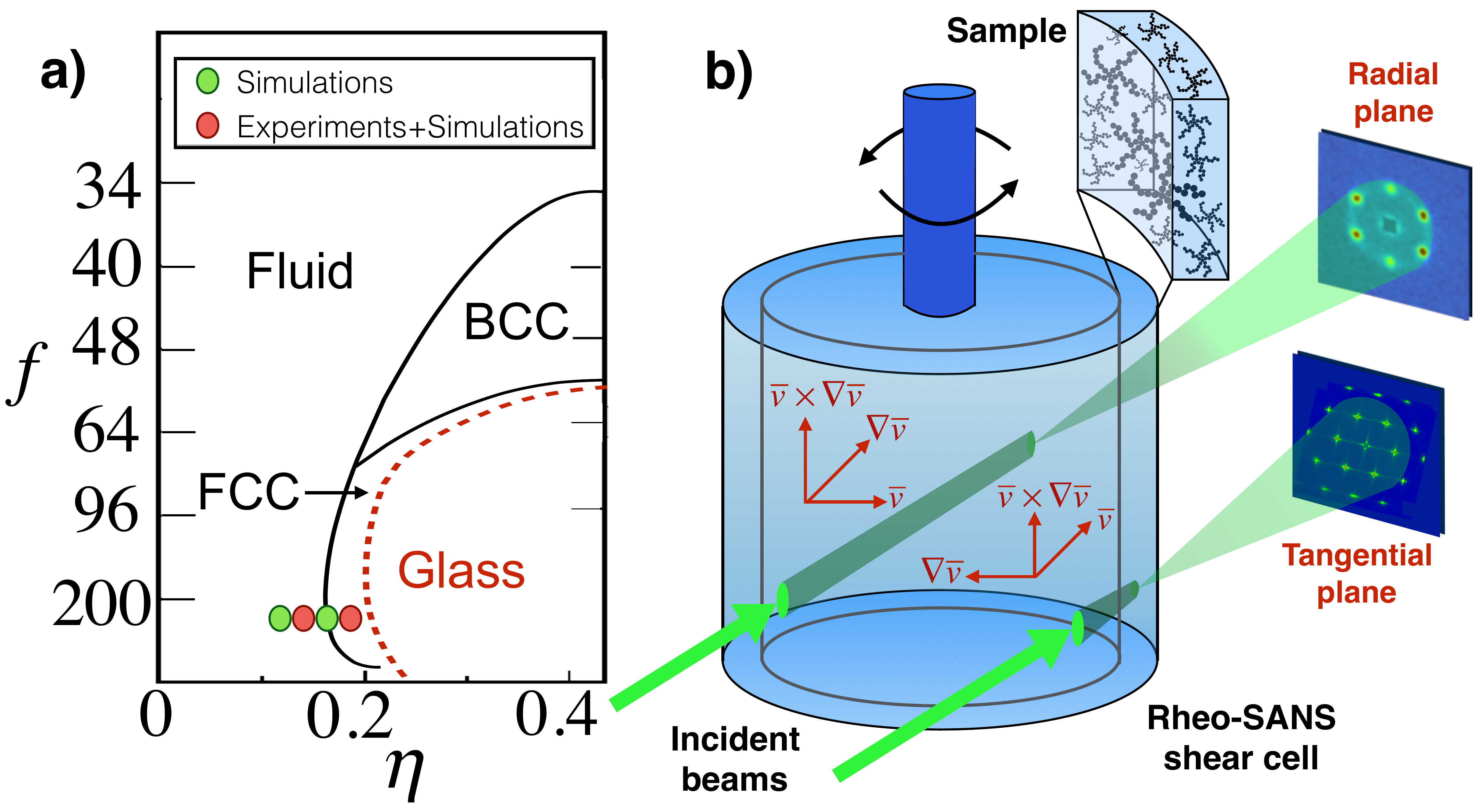}
\caption{a) Theoretical state diagram of star polymers\protect\cite{watzlawek1999phase,foffi2003structural} in the $(f,\eta)$ plane. State points investigated in this work are marked with symbols. 
(b) Schematic illustration of the rheo-SANS experiments and measured diffraction patterns in the radial (experimental data) and tangential (simulations data) plane. See text for details.}
\label{fig1}
\end{figure}

We investigate 1,4-polybutadiene stars with functionality $f = 203$ arms and arm molar mass of 30500 g/mol\cite{gauthier2010synthesis}. The hydrodynamic radius in toluene is 45 nm and the overlap concentration $c^*=27$ mg/ml. The softness of the stars can be quantified by the softness parameter $SP=0.11$\cite{vlassopoulos2014tunable,daoud}, as described in the Supplementary Information (SI). We study different concentrations ($2c^*-2.2c^*$), corresponding to a range of packing fractions $\eta\approx0.15-0.167$ \cite{watzlawek1999phase}, in the vicinity of the mestastable glassy regime shown in the phase diagram of Fig~\ref{fig1}(a). The samples do not crystallize in the absence of external field for the investigated time (1 day). The rheological characterization was performed by means of dynamic oscillatory measurements using a sensitive stress-controlled rheometer operating in the strain-controlled mode (see Fig. S1).  Rheo-SANS measurements were carried out at the Swiss spallation neutron source (SINQ) of the Paul Scherrer Institut in Villigen, Switzerland. The rheo-SANS setup combined SANS and a stress-controlled rheometer which offered the possibility of performing measurements in the radial (velocity-vorticity, $\vec{v},\vec{v}\times\nabla\vec{v}$) and tangential (velocity gradient-vorticity, $\nabla\vec{v},\vec{v}\times\nabla\vec{v}$) planes, as illustrated in Fig~\ref{fig1}(b). Further details are provided in the SI.

The experimental investigations are complemented by numerical simulations of particles interacting via a coarse-grained, ultrasoft effective potential 
which mimics the interactions between star polymers\cite{likos1998star,likos2001effective}. We perform molecular dynamics (MD) simulations for $N=2000$ stars with functionality $f=203$
 at different packing fractions (Fig.~\ref{fig1}(a)).  We use a steady shear protocol at fixed shear rate $\dot{\gamma}$ complemented by Lees-Edwards boundary conditions\cite{lees1972computer} and 
a dissipative particle dynamics (DPD) thermostat\cite{zausch2008equilibrium,nicolas2016effects}. 
To quantify crystallization, we calculate local and averaged bond order parameter distributions\cite{steinhardt1983bond,lechner2008accurate}, assigning solid-like nature to each particle and also distinguishing between different crystal structures\cite{russo2012microscopic}. We also monitor the fraction of solid-like particles and define a crystallization time $t_X$ when this fraction reaches 20\%\cite{valeriani2012compact}. Numerical results are averaged over five independent realizations.

To compare experimental and numerical results obtained under different shear protocols, we use the P\'{e}clet number $Pe=\dot{\gamma}\tau_{B}$, where 
$\tau_{B}$ is the Brownian time defined in terms of the self-diffusion coefficient at infinite dilution. With this definition P\'{e}clet numbers vary in the range $10^{-5}\lesssim Pe \lesssim 10^{-1}$ (see SI). The experimental shear rate is $\dot{\gamma}=\gamma_0\omega$, with $\gamma_0$ the strain amplitude and $\omega$ the frequency. In both experiments and simulations, we also calculate the degree of order parameter (DOO), 
which captures the increase of the intensity in the diffraction patterns associated to the growth of crystalline order in the system. More details are provided in the SI.

\begin{figure}
\includegraphics[width=8.5 cm]{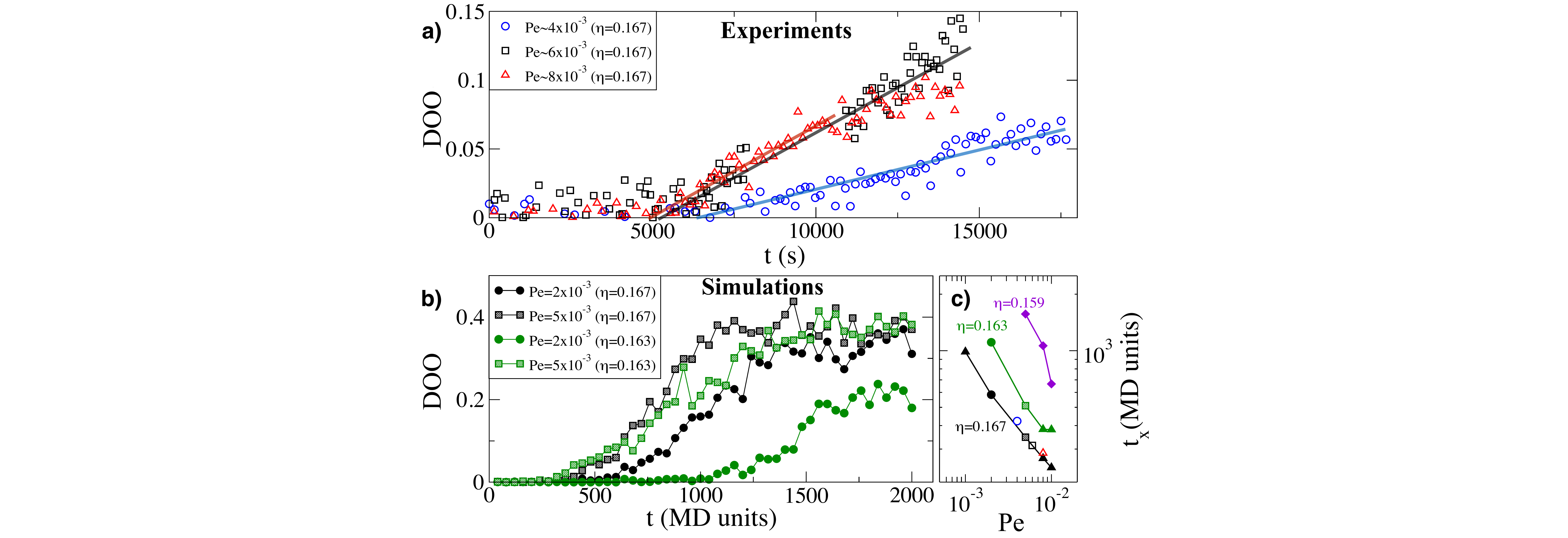}
\caption{Degree of ordering versus time calculated for experiments (a) and simulations (b). 
In (a) the shear parameters are: $\omega=10$ rad/s and $\gamma_{0}=11.5\%$ (squares), $\omega=5$ rad/s and $\gamma_{0}=15\%$ (circles), $\omega=10$ rad/s and $\gamma_{0}=15\%$ (triangles).  
Lines are guides to the eye to highlight the onset of crystallization;
(c) nucleation time $t_X$ as function of $Pe$ for simulations at different packing fractions (filled symbols) and for experiments (scaled by an arbitrary factor) at $\eta=0.167$ (open symbols as in panel (a)).
}
\label{Fig2}
\end{figure}

For the investigated packing fractions, the system at rest is a metastable liquid or glass as revealed by linear viscoelastic measurements (see Fig. S1), reflecting the proximity of the studied state points to the fluid-crystal (fcc) boundary predicted theoretically\cite{watzlawek1999phase,stiakakis2010slow}. 
In all cases the samples were sheared at rates corresponding to the solid-like region of the linear viscoelastic spectrum (Fig. S1). To monitor the crystallization process, we report in Fig.~\ref{Fig2} the DOO for the amorphous (fluid or glass) to crystal transition observed in (a) experiments and (b) simulations, showing 
the same qualitative trends: (i) there is an induction time for crystallization to occur; (ii) at the same star packing fraction crystallization is faster and more pronounced with increasing 
$Pe$;
(iii) under the same shear conditions, an increase of $\eta$ facilitates and speeds up crystallization. These features are also evident in Fig.~\ref{Fig2}(c), where the crystallization time $t_X$ is reported as a function of $Pe$ for different values of $\eta$. It is also found experimentally that frequency has a stronger influence on the DOO (Fig.~\ref{Fig2}(a)), and thus on the nucleation time, with respect to strain amplitude. These results confirm earlier results for 
hard sphere systems\cite{ackerson1988shear,ackerson1990shear,blaak2004crystal,holmqvist2005crystallization,richard2015role}, suggesting that in general large enough shear rates
are needed in order to induce crystallization. Some quantitative differences between simulations and experiments (the former being more sensitive to shear rate) are attributed to the different protocols used. 

The calculation of bond order parameters\cite{lechner2008accurate,russo2012microscopic} in simulations reveals that the fluid-to-crystal transition in most cases, and always for large enough $Pe$, gives rise to a fcc-like crystal (see Fig.~S2). While for very low values of $Pe$ ($Pe \lesssim 10^{-3}$) no crystallization takes place, for intermediate values of $\dot{\gamma}$ we observe a two-step process: at first a transition occurs from fluid to a bcc-like crystal, later followed by a second transition to a fcc/hcp-like crystal. 
Both transitions are accompanied by clear discontinuities in the energy of the system (Fig.~S3). 
A crystal-to-crystal transition is found for $0.159\leq\eta\leq0.167$ at sufficiently small $Pe$. On decreasing packing fraction, the crystal-to-crystal transition is observed by increasing $Pe$. 

\begin{figure*}
\centering \includegraphics[width=18cm]{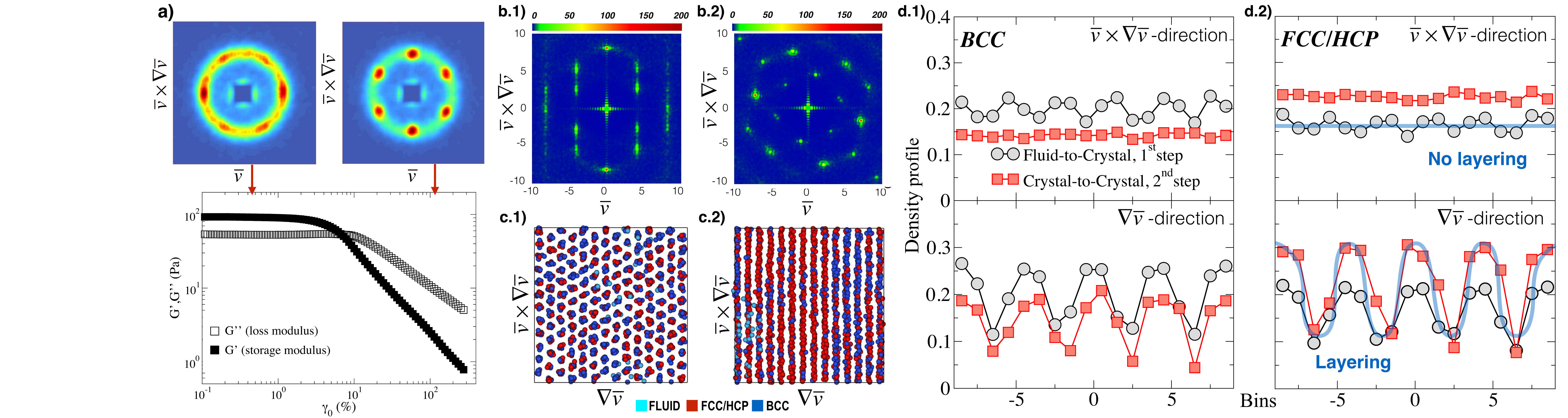}
\caption{(a) SANS diffraction patterns in the radial direction and strain amplitude sweep depicting the measured $G'$ and $G"$ for stars at $\eta=0.167$. The flipping between the two crystals was observed for $\omega=5$ rad/s in a range of strain amplitude (0.1-300 $\%$), amounting to a range in P\'{e}clet number ($0-8\times10^{-2}$). Vertical arrows indicate the strain values at which the images were extracted; 
(b-d) Numerical results for simulations at $\eta=0.167$ and $Pe=2\times10^{-3}$: radial diffraction patterns after the $1^{st}$ step (b.1) and after the $2^{nd}$ step (b.2); system snapshots in the tangential view after $1^{st}$ step (c.1) and after $2^{nd}$ step (c.2). Here the size of particles is reduced to help visualization;
density profiles measured for bcc (d.1) and fcc/hcp (d.2) particles in the vorticity direction (upper panels) and gradient direction (lower panels). Continuous lines are drawn to indicate the behaviour of the density profiles in the absence (flat line) or in the presence (periodic curve) of layers of particles.}
\label{Panel}
\end{figure*}

Such a behavior is also found in experiments at $\eta=0.167$ upon the application of strain amplitude from $0.1\%$ to $300\%$ (within 600s) with a frequency $\omega=5$ rad/s, as reported in Fig.~\ref{Panel}(a): a transition from amorphous glass to crystal takes place at $Pe\sim 1.4 \times 10^{-4}$ ($\gamma_{0}=0.5\%$), followed by a crystal-to-crystal transition at strain amplitudes higher than $120\%$ ($Pe\gtrsim 3.3 \times10^{-2}$), well above rheological yielding. A crystal-to-crystal transition was only observed for $\omega=5$ rad/s and not for larger frequencies, suggesting that not too high shear rates are required to induce the first transition to an intermediate crystal structure. Although  it is not straightforward to compare parameters obtained with different shear protocols, these findings are in qualitative agreement with simulations.

From the radial rheo-SANS diffraction patterns shown in Fig.~\ref{Panel}(a) 
we can speculate that a transition takes place between two hexagonal order structures oriented along different directions. To verify this interpretation, we rely on numerical simulations and calculate diffraction patterns from the particle coordinates\cite{metere2016smectic}. In Fig.~\ref{Panel}(b.1,b.2) we report the diffraction patterns in the radial direction of the first and second crystal respectively. The numerical results are again in good agreement with the experimental SANS patterns (Fig~\ref{Panel}(a)),
despite the difference in the used shear protocol. 
To visualize the two (fluid-to-crystal and crystal-to-crystal) transitions, movies from the simulations are presented in the SI, while snapshots of the two crystal structures in the tangential plane are reported in Fig.~\ref{Panel}(c.1-c2). After completing the $1^{st}$ step (Fig.~\ref{Panel}(b.1)), the crystal is organized into two different layers oriented orthogonally both to the vorticity and to the velocity gradient directions due to the bcc geometry, while after the $2^{nd}$ step the layers reorganize and become orthogonal with respect to the gradient direction only (Fig.~\ref{Panel}(b.2)). 
These features are clearly identified by looking at the calculated density profiles along different directions respectively for bcc (Fig.~\ref{Panel}(d.1)) and fcc/hcp particles (Fig.~\ref{Panel}(d.2)).  
We observe oscillations in the density in both the velocity-gradient and vorticity directions after the first step.
However, after the $2^{nd}$ step, a flat profile is observed for the vorticity axis, while oscillations survive in the $\nabla\vec{v}$ direction. The layers of fcc particles are only orthogonal to $\nabla\vec{v}$ at all times.
Figure~\ref{Panel}(d.2) also shows that an enhancement of oscillations along the $\nabla\vec{v}$ axis after the $2^{nd}$ step for fcc particles is associated to a decrease of the same oscillation for bcc ones. 
These features clearly indicate that the formation of a bcc lattice is responsible for the peculiar structure observed in the $1^{st}$ step. Indeed, the layering orthogonal to the vorticity is completely lost once these particles reorganize into a fcc, giving rise to the layers commonly observed in other shear-induced experiments\cite{eiser2000nonhomogeneous,blaak2004crystal,lopez2015layering,PhysRevFluids.2.093301}.

To connect our findings with previous observations of crystal-to-crystal transformations, 
we investigate whether in our system there is evidence of intermediate melting, at least locally, which could help the (re)organization into a different lattice. To this aim we monitor
the fraction of particles of each species (liquid, fcc, bcc, hcp) during the second step, finding that melting does not occur during the bcc-like to fcc-like transition (see Fig.~S4). This constitutes a striking difference with respect to the case of thermoresponsive microgels studied in Ref. \cite{peng2015two} and may be attributed in part to the different protocol used in that work, where the transition was induced by varying the temperature, rather than by shear.
On the other hand, Refs.\cite{lopez2015layering,chu2015colloidal} reported intermediate melting in the presence of shear, without notable soft particle deformation.
For the star polymers under shear studied in the present work, 
the reorganization of the crystal lattice between two competing structures occurs without intermediate melting even at the local level. Instead, we observe a sudden change, i.e., a 'flipping', between bcc and fcc lattice, which provides an alternative mechanism to realize a crystal-to-crystal transition in this system.  These findings are linked to the peculiar 
nature of star polymers which allows for a direct transformation between two crystals, thanks to their ultrasoft interactions.  
Indeed, according to the theoretical phase diagram (Fig.~\ref{fig1}(a)), for the studied state points the system is approaching a glass transition but its underlying equilibrium state is the fcc crystal. Being dominated by Yukawa-like repulsions at low packing fractions, the free energy difference between fcc and bcc structure is very small\cite{robbins1988phase}.  Thus, the competition between these two crystalline structures, which is influenced by $Pe$, determines the final state of the sheared system.  Based on our results we suggest that at high enough $Pe$ the system experiences a  fluid-to-crystal directly into fcc due to the large rearrangements induced by shear. On the other hand, at lower $Pe$ shearing is not strong enough and the system is only able to complete the crystallization process in two steps, by first attaining an intermediate (metastable) bcc-like state and then reaching a situation comprising a mixture of bcc and fcc structures. This is confirmed by the fact that at even lower $Pe$ crystallization is not observed, whereas the value of threshold $Pe$ to achieve a two-step crystallization increases with decreasing packing fraction. Importantly, once the $2^{nd}$ step is reached, the crystal does not melt upon shear cessation, but remains stable over time in both experiments and simulations. We stress that the final structure with layers parallel to the flow is in agreement with previous studies of shear-induced crystallization\cite{eiser2000nonhomogeneous,blaak2004crystal,lopez2015layering,PhysRevFluids.2.093301}.  However, the intermediate structure occurring after the $1^{st}$ step and the mechanism behind the crystal-to-crystal transition are novel features of the present study, that are attribute to to the ultrasoftness of colloidal star polymers.

In summary, the application of shear induces crystallization of ultrasoft star polymer suspensions at packing fractions in the vicinity of the glass line. 
The good agreement between experiments and simulations despite the different shear protocol used strongly supports the generality of the results.
In most cases, a fluid-to-crystal transition under shear is found, which is facilitated by increasing $Pe$ and increasing packing fraction. 
However, for $0.159\leq\eta\leq0.167$ there exists an intermediate range of Pe where stars undergo a distinct crystal-to-crystal transition. 
The transition consists of a transformation between a bcc-dominated and an fcc-dominated crystal, 
which occurs via a flipping of the crystal structure and not by an intermediate melting, differently from previous studies. 
Our results indicate that the combination of shear and softness is important for shedding light on the fundamental physics underlying phase transitions as well as tailoring the organization of soft materials with desired properties. To this end, the tunable softness of star polymers is very valuable and future directions will include the control and manipulation of crystal-to-crystal transitions in different regions of the phase diagram, changing both functionality and packing fractions.


JRF and JM equally contributed to this work. We are grateful to E. Stiakakis for help with samples preparation and characterization and C.N. Likos for enlightening discussions. JRF, DV, EZ acknowledge support from ETN-COLLDENSE (H2020-MCSA-ITN-2014, Grant No. 642774); JM, DV from the Greek General Secretariat for Research and Technology in the framework of the program Thalis (project METAASSEMBLY); DV from the Aage og Johanne Louis-Hansen Foundation; MG from the Natural Sciences and Engineering Research Council of Canada; NG, EZ from ERC Consolidator Grant 681597 MIMIC.

\bibliography{biblio}

\end{document}


\title{Supplementary material for Crystal-to-crystal transition of star colloids under shear}

\author{J. Ruiz-Franco$^{1}$, J. Marakis$^{2,3}$, N. Gnan$^{1,4}$, J. Kohlbrecher$^5$, M. Gauthier$^6$, M. P. Lettinga$^{7,8}$, D. Vlassopoulos$^{2,3}$, E. Zaccarelli$^{1,4}$}
\address{$^1$Dip. Di Fisica, Sapienza Universit\`a di Roma, P.le A. Moro 5, 00185 Rome, Italy}
\address{$^2$Foundation for Research and Technology-Hellas (FORTH), Institute of Electronic Structure and Laser (IESL), 100 N. Plastira Str., GR-70013 Heraklion, Greece}
\address{ $^3$University of Crete, Department of Materials Science and Technology, P.O.Box 2208, GR-71003 Heraklion, Greece}
\address{ $^4$CNR Institute of Complex Systems, Uos Sapienza, Rome, Italy}
\address{$^5$ Laboratory for Neutron Scattering, ETH Zurich \& Paul Scherrer Institut, 5232 Villigen PSI, Switzerland}
\address{$^6$ University of Waterloo, Department of Chemistry, Waterloo, ON N2L 3G1, Canada}
\address{$^7$Laboratory for Soft Matter and Biophysics, KU Leuven, Celestijnenlaan 200D, B-3001 Leuven, Belgium}
\address{$^8$Institute of Complex Systems (ICS-3), Forschungszentrum J\"ulich, 52425 J\"ulich, Germany.}
\date{\today}

\date{\today}

\maketitle

\centerline{{\bf Additional details on experimental methods}}

{\it Star softness:} 
To quantify the softness of the stars we make use of the Daoud-Cotton model\cite{daoud}, which can be applied because the stars fullfill the condition 
$N >> f^{1/2}v_{m}^{-2}$, where $v_{m}$ is the  excluded volume of the monomer. The softness parameter is defined as $SP=\frac{R_{cor}}{R_{sw}}=f^{3/10}N^{-3/5}$, where $R_{sw}$ and $R_{cor}$ are the radii of the swollen regime and of the core, respectively. This yields a rather low value $SP=0.11$, indicative of rather soft objects\cite{daoud}. 

\begin{figure}
\centering \includegraphics[width=9 cm]{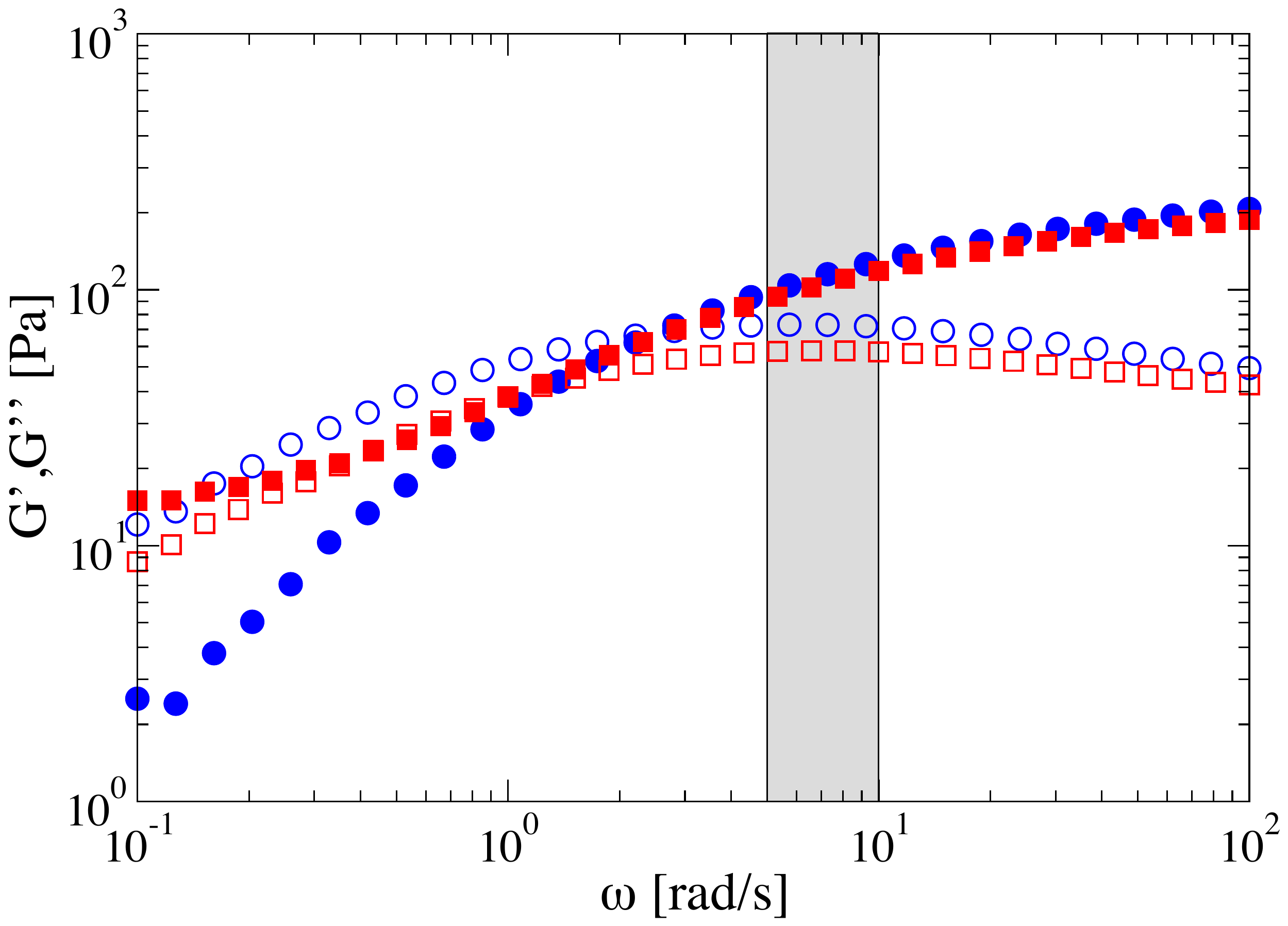}
\caption{Dynamic storage (G', full symbols) and loss moduli (G'',open symbols) of the stars measured with $\gamma_{0}=0.01\%$,
at $c/c^*=2.05$ corresponding to $\eta=0.152$ (blue circles) , and at $c/c^*=2.18$ corresponding to
$\eta=0.167$ (red squares). The shaded region indicates the regime of frequencies used in rheo-SANS ($5 - 10$ rad/s).
}
\label{fig:SI-G}
\end{figure}

{\it Rheological characterization:} 
Shear rheometric measurements were performed with a sensitive stress-controlled rheometer operating in the strain-controlled mode (Physica MCR 501, Anton Paar, Austria). A coaxial cylindrical Couette geometry with an inner rotating titanium bob of diameter 49 mm and an outer glass cup of diameter 50 mm was used. Their length was 120 mm. The temperature was set at $20^{\circ}C$ by means of a water-ethylene glycol recirculating bath. The outer atmosphere was saturated with toluene by means of soaked tissues in order to minimize the risk of toluene evaporation for about 1 hour. 
Measurements included (i) dynamic frequency sweeps by imposing a linear oscillatory strain $\gamma=\gamma_0 sin(\omega t)$ in order to probe the viscoelastic relaxation spectrum, i.e., the frequency-dependent storage ($G'$) and loss ($G''$) moduli in the range $0.1-100$ rad/s. The stress response is $\sigma=\sigma_0 sin(\omega t) +\delta=\gamma_0 (G' sin(\omega t) + G'' cos(\omega t))$ with $\sigma_0$ being the stress amplitude and $\delta$ the phase angle ($tan\delta=G''/G'$). 
The viscoelastic spectra of the studied concentrations are shown in Fig.~\ref{fig:SI-G}; (ii) dynamic strain sweeps with a duration of about 8 min, in order to determine the linear viscoelastic and yielding regimes. It involves oscillations at constant frequency $\omega$ and continuously increasing $\gamma_0$ from $0.1\%$ to $300\%$ in strain amplitude. The experimental Brownian time, defined as $\tau_B=R_H^2/D_0$, where $R_H$ is the hydrodynamic radius and $D_0$ the self-diffusion coefficient of the stars at infinite dilution, is $\tau_B=5.485$~ms, so that the range of explored P\'{e}clet numbers $Pe=\dot{\gamma} \tau_B$ is $2.7\times 10^{-5} < Pe < 8.2\times 10^{-2}$ (see comments in the main text and in the P\'{e}clet number subsection below about $Pe$ values); (iii) dynamic time sweeps at different frequencies and strain amplitudes.  These tests were preceded by steady shear measurements at different rates (from $1s^{-1}$ to $100s^{-1}$) in order to shear-melt (rejuvenation process) the structure of the system. This protocol allowed erasing the sample's history (possible residual stresses during loading), ensuring reproducible initial conditions for the measurements.

{\it Rheo-SANS:} The same rheometer was used for the SANS measurements under flow. The Couette geometry allowed performing measurements in both radial (velocity-vorticity) and tangential (velocity gradient-vorticity) planes by sending the neutron beam along the velocity gradient and velocity directions, respectively (see Fig.1a of the manuscript). From the Rheo-SANS measurements, we calculate the degree of order defined as $DOO=\frac{\sum_{i=1}^{6}Ip_{i}-\sum_{i=1}^{6}Iv_{i}}{\sum_{i=1}^{6}Ip_{i}}$, where $Ip$ and $Iv$ are the intensity of the peaks and valleys of the diffraction patterns.

\centerline{{\bf Additional details on numerical methods}}

{\it Simulations:} 
In order to mimic the effect of the solvent acting on the colloids and to ensure Galileian invariance, we use a dissipative particle dynamics (DPD) thermostat\cite{zausch2008equilibrium,nicolas2016effects} coupled to our equations of motion, i.e. 
\begin{equation}
	\label{eq:DPD}
	\begin{array}{c}
	\frac{d\mathbf{r}_{i}}{dt}=\mathbf{v}_{i}\\
	m_{i}\frac{d\mathbf{r}_{i}}{dt}=\sum_{j\ne i}\left(\mathbf{F}_{ij}^{C}+\mathbf{F}_{ij}^{D}+\mathbf{F}_{ij}^{R}\right)
	\end{array}
\end{equation}
Here, $\mathbf{F}_{ij}^{C}$ are the conservative forces, associated to the interaction potential
between particles $i$ and $j$, while $\mathbf{F}_{ij}^{D}$ and $\mathbf{F}_{ij}^{R}$ are a dissipative and random force,
respectively,
\begin{eqnarray}
  \label{eq:Motion}
      \mathbf{F}_{ij}^{D}&=&-\xi\omega^{D}\left(r_{ij}\right)\left(\hat{\mathbf{r}}_{ij}\cdot\mathbf{v}_{ij}\right)\hat{\mathbf{r}}_{ij}\nonumber\\
      \mathbf{F}_{ij}^{R}&=&\sigma\omega^{R}\left(r_{ij}\right)\theta_{ij}\hat{\mathbf{r}}_{ij}
\end{eqnarray}
with $\xi$ a friction coefficient, $\theta_{ij}=\theta_{ji}$ uniform random numbers with zero
mean and unit variance, $\mathbf{v}_{ij}=\mathbf{v}_{i}-\mathbf{v}_{j}$ the relative velocity
between particle $i$ and $j$, $\hat{\mathbf{r}}_{ij}$ the unit vector of the vector
$\mathbf{r}_{ij}=\mathbf{r}_{i}-\mathbf{r}_{j}$ and $r_{ij}=\left|\mathbf{r}_{i}-\mathbf{r}_{j}\right|$
the distance between particle $i$ and $j$.  The two weight functions are related as
$w^{D}\left(\mathbf{r}_{ij}\right)=\left[w^{R}\left(\mathbf{r}_{ij}\right)\right]^{2}=\left(1-\frac{\mathbf{r}_{ij}}{r_{c}^{DPD}}\right)^{s}$
if $r<r_{c}^{DPD}$, where $r_{c}^{DPD}$ is a cutoff value that we have fixed to be equal to the first minimum 
of the radial distribution function of the system in equilibrium. 
The equations of motion were computed with the scheme of Peters \cite{peters2004elimination} using a time step of $\Delta t=0.002$.
We fix $\xi=25$ and $s=1$.

{\it Effective potential:}. 
Star polymers are modeled by the effective center-center potential developed by Likos and coworkers \cite{likos1998star}:
\begin{equation}
  \label{eq:Pot}
  \beta V\left(r\right)=\left\{ \begin{array}{c}
  \frac{5}{18}f^{\frac{3}{2}}\left[-ln\left(\frac{r}{\sigma}\right)+\left(1+\frac{\sqrt{f}}{2}\right)^{-1}\right]\  for\  r\leq\sigma\\
  \frac{5}{18}f^{\frac{3}{2}}\left(1+\frac{\sqrt{f}}{2}\right)^{1}\left(\frac{\sigma}{r}\right)exp\left[-\frac{\sqrt{f}\left(r-\sigma\right)}{2\sigma}\right]\ for \ r>\sigma
  \end{array}\right. 
\end{equation}
where $\sigma$ is the corona diameter and $f$ is the functionality (number of arms). In the simulations, we consider
monodisperse stars. The potential cut-off is fixed to $r_{c}=3.0\sigma$. 

{\it Simulation units:} 
Length is measured in units of $\sigma$ while energy in units of $k_BT$.  Simulations are performed at $k_BT=1$ due to the athermal nature of the effective potential for all studied packing fractions  $\eta=\frac{\pi\sigma^{3}}{6}\frac{N}{V}$, where $N=2000$ is the number of particles and $V$ the volume of the simulation box.  Time is measured in units of $\sqrt{\frac{m\sigma^{2}}{\varepsilon}}$. 

{\it P\'{e}clet number:}
In order to compare our results with experiments we also use the P\'{e}clet number $Pe=\dot{\gamma}\tau_{B}$ evaluated from the self-diffusion coefficient at infinite dilution. 
In our simulations, $\tau_B\sim10^{-2}$ MD units, so that the explored range of P\'{e}clet numbers is $10^{-4}< Pe < 10^{-2}$. This range of values is comparable to the experimental ones, despite the difference in the used shear protocol and to the absence of explicit solvent in the simulations.   The  P\'{e}clet numbers used in our investigations seem to be very low because they are calculated with respect to $D_0$, but when the self-diffusion coefficient $D$ at the studied packing fractions is taken into account, the so-called dressed P\'{e}clet number $\tilde{Pe}=Pe D_0/D$ is much larger, i.e.  $1\lesssim \tilde{Pe}\lesssim 10^{2}$ for $\eta=0.167$. These numbers imply quite a large shear rate imposed on the system. Since in experiments we do not have access to $D$ but only to $D_0$, we use the standard P\'{e}clet in order to have a meaningful comparison with simulations.

{\it DOO calculation in simulations:} 
Given a diffraction pattern, we define the degree of order as 
\begin{equation}
DOO=\frac{\sum_{q_{\vec{v}},q_{\vec{v}\times \nabla \vec{v}}} I_p(q_{\vec{v}},q_{\vec{v}\times \nabla \vec{v}})-I_v(q_{\vec{v}},q_{\vec{v}\times \nabla \vec{v}})}{\sum_{q_{\vec{v}},q_{\vec{v}\times \nabla \vec{v}}} I_p(q_{\vec{v}},q_{\vec{v}\times \nabla \vec{v}})}
\end{equation} 
\noindent where $(q_{\vec{v}},q_{\vec{v}\times \nabla \vec{v}})$ are the vectors in the reciprocal space corresponding to the radial direction, $I_p(q_{\vec{v}},q_{\vec{v}\times \nabla \vec{v}})$ is the intensity value for each set of wave vectors (peaks and background), while  $I_v(q_{\vec{v}},q_{\vec{v}\times \nabla \vec{v}}) < I_{threshold}$ are all the 
 intensities with a value smaller than the threshold value $I_{threshold}$ that we set to $I_{threshold}=20$. 
 In such a way, when no clear signal of crystallization is found (i.e. all the intensities are smaller than $I_{threshold}$), $I_p(q_{\vec{v}},q_{\vec{v}\times \nabla \vec{v}})=I_v(q_{\vec{v}},q_{\vec{v}\times \nabla \vec{v}})$ and $DOO=0$. On the other hand when the intensity peaks signal the increasing order in the system  $I_p(q_{\vec{v}},q_{\vec{v}\times \nabla \vec{v}})>I_v(q_{\vec{v}},q_{\vec{v}\times \nabla \vec{v}})$ and $DOO>0$.

\begin{figure}
\centering \includegraphics[width=7.75 cm]{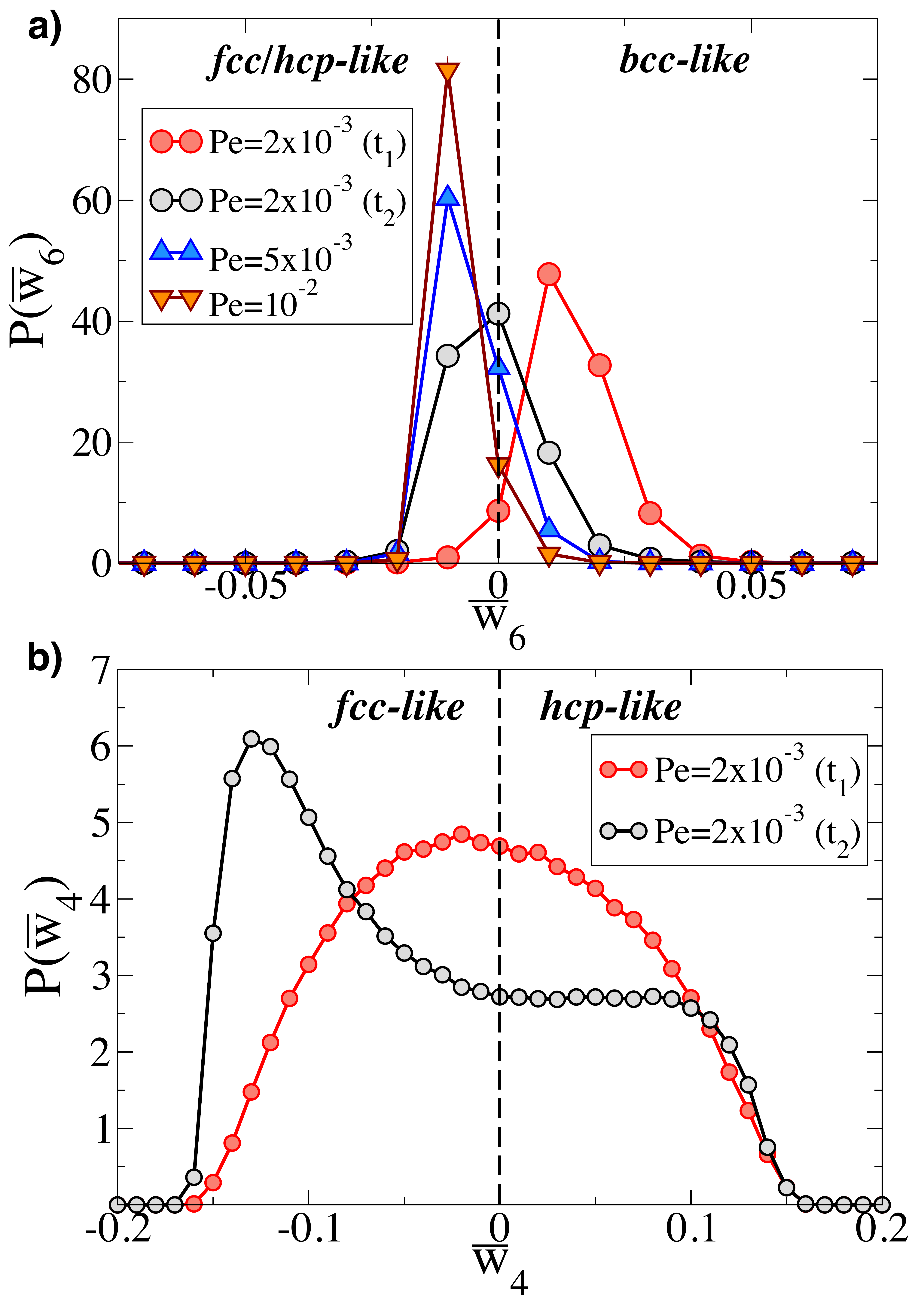}
\caption{Simulation results for $\eta=0.167$: 
(a) distribution of the orientational order parameters $\overline{w}_{6}$ for different P\'{e}clet numbers. Here, $t_{1}$ and $t_{2}$ are two time values referring to the fluid-to-crystal and crystal-to-crystal transitions respectively (i.e. $t_{1} < t_{2}$) for $Pe=2\times10^{-3}$; 
(b) distribution of the orientational order parameters $\overline{w}_{4}$ for $Pe=2\times 10^{-3}$ at times $t_{1}$ and $t_{2}$ .}
\label{fig:SI2_V}
\end{figure} 
 
{\it Density profiles:}  
The density profile is computed by dividing the simulation volume along a given axis into boxes of thickness $\sigma$ and area $L\times L$ where $L$ is the edge of the cubic simulation box. We then evaluate the density of particles within each box defined as $N_{box}/V_{box}$ where $N_{box}$ is the number of particles and $V_{box}$ is the volume of a box of thickness $\sigma$. The evolution of the density along a chosen axis gives information on the layering effect occurring when crystals are formed. In particular a flat profile indicates that the distribution of particles is homogeneous, while oscillations highlight the formation of layers orthogonal to the axis along which the density profile has been evaluated.
 
 \centerline{{\bf Additional results}}
 
{\it Analysis of the crystal order:}
The analysis of the crystal structure is based on the calculations of the bond orientational order parameter distributions\cite{steinhardt1983bond,lechner2008accurate}. In particular, the local observable $\overline{q}_{6}$ is used to assign solid-like nature to each particle in the simulation according to the number of solid-like connections as done in \cite{pusey2009hard}. 
Following the work of Russo and Tanaka\cite{russo2012microscopic}, to distinguish between different crystal structures, we calculate $\overline{q}_{6}$ to differentiate liquid-like and solid-like particles. In addition to separate within solid particles, i.e. bcc-like,fcc-like and hcp-like we also calculate the averaged orientational order parameters  $\overline{w}_{4}$ and $\overline{w}_{6}$. In Fig.~\ref{fig:SI2_V}(a) the distribution of $\overline{w}_{6}$ is reported for several values of the P\'{e}clet  number. At high Pe, the fluid undergoes a single transition to a fcc-like crystal. 

For $Pe=2\times 10^{-3}$ we calculate $P\left(\overline{w}_{4}\right)$ at time $t_{1}$ (after the fluid-to-crystal transition) and $t_{2}$ (after the crystal-to-crystal transition), finding that the crystalline order at $t_{1}$ is bcc-like, later becoming fcc-like at $t_{2}$. In Fig.~\ref{fig:SI2_V}(b) the distribution of $\overline{w}_{4}$  for $t_1$ and $t_2$ is reported, allowing to characterize the fcc or hcp order. We find that  while in the first step both types of order are present, in the second crystal the fcc-like nature is dominant over the hcp one.

\begin{figure}
\centering \includegraphics[width=8 cm]{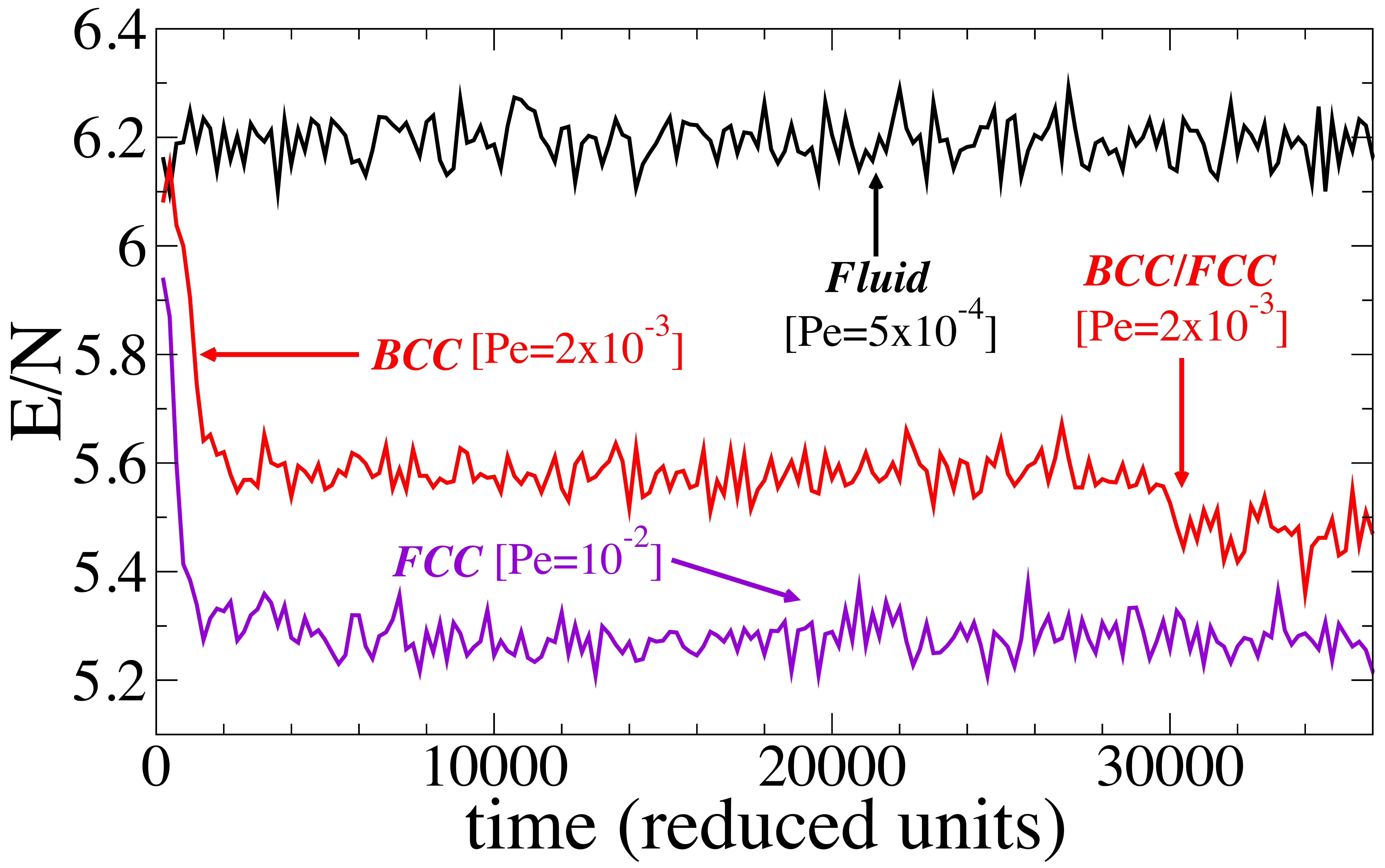}
\caption{Potential energy per particle versus time for $\eta=0.167$ and different values of the P\'{e}clet number.}
\label{fig:SI3}
\end{figure}


{\it The potential energy behaviour:} 
In Fig. ~\ref{fig:SI3} the potential energy per particle versus time is reported for simulations at $\eta=0.167$  and different P\'{e}clet numbers. While for $Pe=5\times 10^{-4}$ the system remains fluid and the energy is constant, for $Pe=10^{-2}$ a sudden drop is observed at early times indicating a fluid-to-crystal transition. However, for $Pe=2\times10^{-3}$, the energy drops twice: a first drop corresponding to a fluid-to-crystal transition at a larger energy value with respect to the previous case, followed at much later times by a second drop. 

\begin{figure}
\centering \includegraphics[width=8.5 cm]{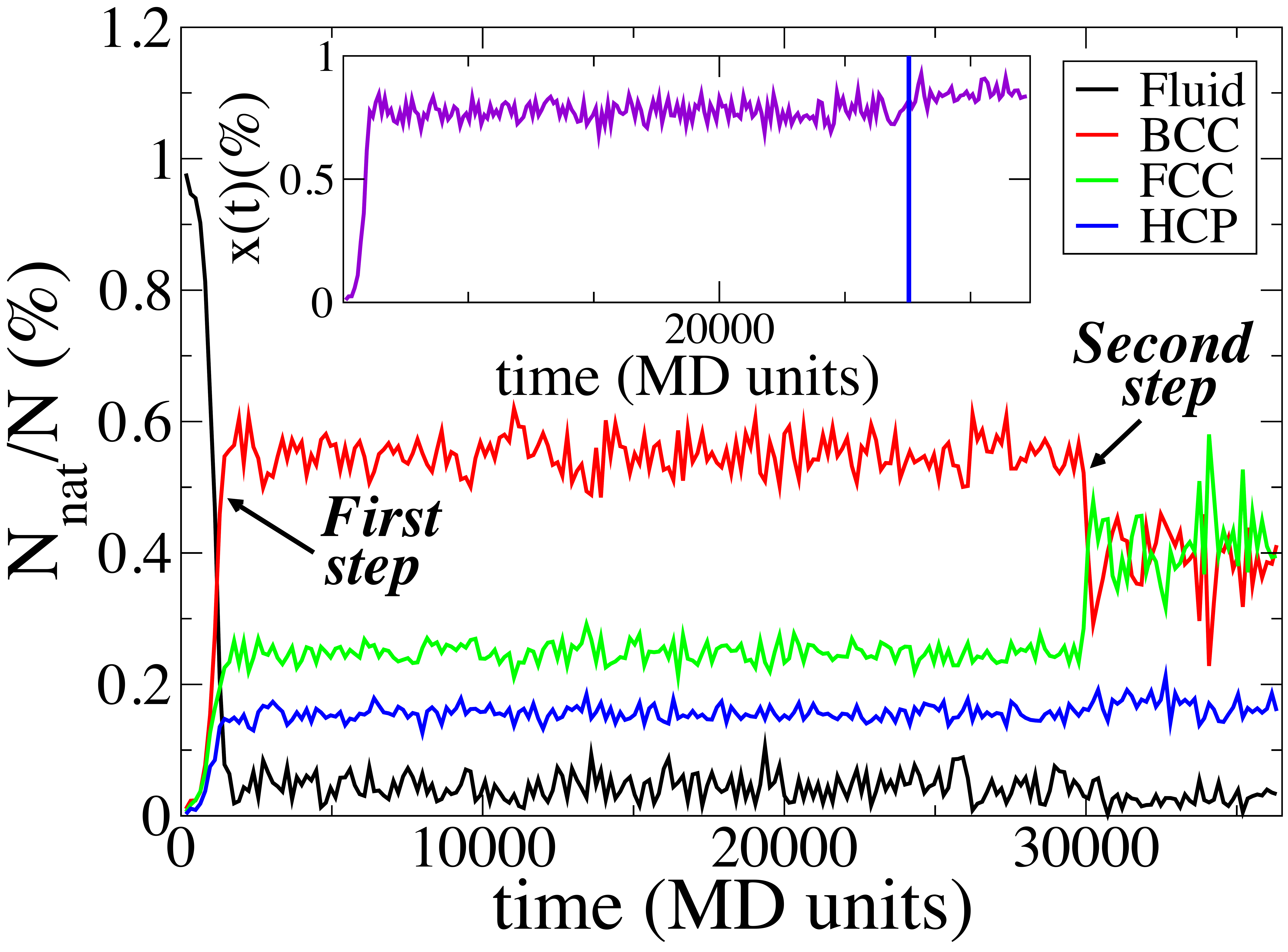}
\caption{Simulation results for $\eta=0.167$: fraction of particles of each species (see legends) versus time. Inset: fraction of solid-like particles $X\left(t\right)$.}
\label{fig:SI4}
\end{figure}

\begin{figure}
\centering \includegraphics[width=8 cm]{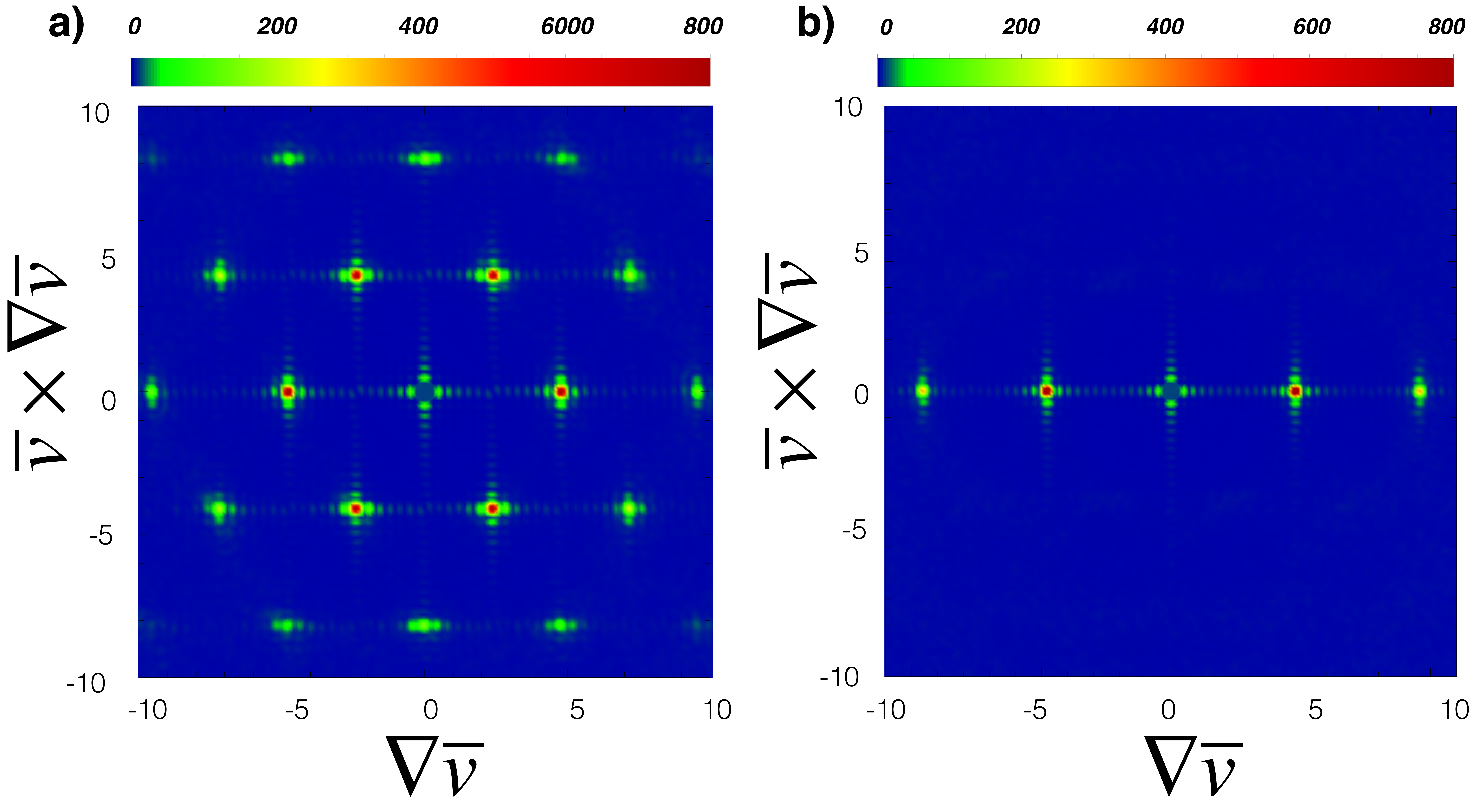}
\caption{Numerical diffraction patterns for $\eta=0.167$ and $Pe=2\times 10^{-3}$ in the tangential direction for (a) $1^{st} step$ and (b) $2^{nd} step$.}
\label{fig:SI5}
\end{figure}

{\it Crystal-to-crystal transition:} 
In Fig. ~\ref{fig:SI4} the fraction of particles of each species, including fluid-like particles, is shown as a function of time. It is clear that in the first step the majority of particles are of bcc type, while in the second step a clear change from bcc to fcc takes place, with the hcp-like particles being constant. Fluid particles remain very few and their fraction does not change across the crystal-to-crystal transition. This is evident also from the inset of Fig. ~\ref{fig:SI4}, where the fraction of particles in the crystalline phase $X\left(t\right)$ is shown to remain constant at all times after the initial fluid-to-crystal transition. Thus, no intermediate melting, even at the local level, occurs across the crystal-to-crystal transition. 

\centerline{{\bf Supplemental Movies}}

In all movies, the size of particles was reduced in order to help visualization.

{\bf Movie S1:} Fluid-to-crystal (fcc) transition at $Pe=10^{-2}$ viewed in the radial plane.

{\bf Movie S2:} Fluid-to-crystal (fcc) transition at $Pe=10^{-2}$ viewed in the tangential plane.

{\bf Movie S3:} Fluid-to-crystal (bcc) transition at $Pe=2\times10^{-3}$ viewed in the tangential plane.

{\bf Movie S4:} Fluid-to-crystal (bcc) transition at $Pe=2\times10^{-3}$ viewed in the radial plane.

{\bf Movie S5:} Crystal (bcc)-to-crystal (fcc) transition at $Pe=2\times10^{-3}$ viewed in the tangential plane.

\bibliography{biblio}